# Safeguarding Smart Inhaler Devices and Patient Privacy in Respiratory Health Monitoring

**Asaju Babajide[1], Almustapha Wakili[1], Michaela Barnett[2], Lucas Potter[3], Xavier-Lewis Palmer[3] and Woosub Jung[1]**

[1]Towson University, USA
[2]Blacks In Cybersecurity Headquarters, Inc. USA
[3]BiosView Labs, Oswego, USA

basaju@towson.edu
awakili@towson.edu
michaela@bichq.org
lpott005@odu.edu
xavierpolymer@gmail.com
woosubjung@towson.edu

**Abstract**: The rapid development of Internet of Things (IoT) technology has significantly impacted various market sectors. According to Li et al (2024), an estimated 75 billion devices will be on the market in 2025. The healthcare industry is a target to improve patient care and ease healthcare provider burdens. Chronic respiratory disease is likely to benefit from their inclusion, with 545 million people worldwide recorded to suffer in patients using these devices can track their dosage, while healthcare providers can improve medication administration and monitor respiratory health (Soriano et al, 2020). The growing prevalence of IoT devices, including intelligent inhalers like the Propeller Health System Smart Inhaler, Breather Fit, and Lookee O2 Ring, underscores the increasing importance of network connectivity and software development. While IoT medical devices offer numerous benefits, they also come with security vulnerabilities that can expose patient data to cyber-attacks. It's crucial to prioritize security measures in developing and deploying IoT medical devices, especially in personalized health monitoring systems for individuals with respiratory conditions. Addressing the security gaps and vulnerabilities in IoT devices is essential to ensure patient data's safety and privacy. Efforts are underway to assess the security risks associated with intelligent inhalers and respiratory medical devices by understanding usability behaviour and technological elements to identify and address vulnerabilities effectively. This work analyses usability behaviour and technical vulnerabilities, emphasizing the confidentiality of information gained from Smart Inhalers. It then extrapolates to interrogate potential vulnerabilities with Implantable Medical Devices (IMDs). Our work explores the tensions in device development through the intersection of IoT technology and respiratory health, particularly in the context of intelligent inhalers and other breathing medical devices, calling for integrating robust security measures into the development and deployment of IoT devices to safeguard patient data and ensure the secure functioning of these critical healthcare technologies.

**Keywords:** Wearable health technology, Cybersecurity threats, Health data privacy, Respiratory monitoring, Cyber risk, Lookee O2

## 1. Introduction

In today's accelerating world, the growth of Internet of Things (IoT) technology emphasizes the vital role of interconnectivity, software development and updates in enhancing security firmware. Experts have projected that by 2030, there will be approximately 50 billion IoT devices connected globally (Tomas, 2019). Healthcare technology is increasingly leveraging IoT connectivity to enable real-time patient monitoring, improve data-driven decision-making and enhance overall care efficiency. Interconnected Healthcare technology faces challenges such as data privacy risks, interoperability issues that can hinder efficient and secure patient care. Additionally, emerging technologies like AI and telemedicine bring new opportunities but also raise concerns around equity, regulation and ethical use. The COVID-19 pandemic has significantly elevated the crucial respiratory health is, especially for people with conditions like asthma and Chronic Obstructive Pulmonary Disease (COPD). With increased public awareness, there is increased focus on on the use of technology in healthcare and corresponding treatment plans for respiratory health. The sensitive nature of vulnerabilities discovered surrounding technologies in this domain makes further development and integration a matter to be taken into more critical analysis. Thus, addressing gaps in digitized health infrastructure development has become most necessary and assessing infrastructure is of utmost importance for patient care.

The persistent nature of known security issues arises from concurrent development with new features or integrations which can cause various risks that significantly augment use/use-cases resulting in the use of more robust security measures (McGraw and Mandl, 2021). Extensive research conducted on the Science Direct and PubMed databases has provided evidence of a notable increase in publications regarding dry powder inhalers (DPIs) in recent years. The study is evidence of bilateral collaboration. This observation further supports the ongoing expansion of interest in inhaled medicine. Concurrently, various retail reports expect a parallel surge in the pulmonary administration route, driven by a rise in the occurrence of respiratory diseases and the necessity for more efficacious combination treatments (Jin et al, 2020). Dry Powder Inhalers (DPIs) are medical devices designed to deliver medication directly to the lungs in the form of a dry powder, often used for conditions like asthma and COPD. They rely on the patient's inhalation effort to disperse medication, making proper usage crucial for effective treatment. The evolving usage of further integration with technology in such devices represents a prime foothold for potential threats, misuses or misconfigurations as we take such devices and integrate them with technology designed to aid users or help monitor for providers.

Digital medical devices are being integrated into patient care protocols because they can monitor, record, and transmit health data in real-time. For example, the Propeller Health System Smart Inhaler helps manage asthma and COPD by monitoring inhaler usage and providing personalized feedback. The Breather Fit is a specialized device for respiratory muscle training, aiming to improve lung capacity and respiratory health. The Propeller Health System Smart Inhaler, Breather Fit, and Lookee O2 Ring are some innovative inventions that help people with respiratory conditions. Regrettably, there has been a lack of emphasis on thoroughly examining or logging related security risks that could compromise patient/user trust. Medical-device manufacturers legally must exercise vigilance and responsiveness towards such security weaknesses or exposures.

**In this paper, we explore the security concerns with those smart inhalers. Our research will be built upon these hypotheses (Murray et al, 2013):**

1. What potential security risks do smart inhaler devices have?
2. How are these medical devices specifically targeted by network attacks?
3. Is there a connection between user behaviour and the existence of device vulnerabilities?

Our work attempts to examine inherent vulnerabilities identified in those devices. This publication presents the initial work, with additional publication(s) in development to follow. Our focus revolves around analysing usability behaviour and technical vulnerabilities, placing a strong emphasis on ensuring the confidentiality of information within Implantable Medical Devices (IMDs).

**Overall, our contributions between this paper and the developing paper are as follows:**

- We address noteworthy privacy and security challenges concerning widespread security vulnerabilities. Additionally, there exist conventional challenges pertaining to the proper distribution of cryptographic keys and the secure yet adaptable control of access.
- We conducted a comprehensive analysis on the significance of implementing robust security protocols to proactively safeguard smart medical devices against potential threats, with the primary aim of guaranteeing the safety and welfare of patients.
- We have presented a summary of the principal issues, thereby stimulating a discourse for future research to confront these obstacles.

## 2. General Overview of Existing Research and Technologies

### 2.1 Vulnerabilities in Medical Devices

Holdsworth et al (2019) emphasizes the comprehensive security landscape of medical devices, underscoring prevalent vulnerabilities and potential implications for well-being. The realm of cybersecurity threats is constantly evolving with rapid changes in methods, attacks and vectors (Hanna, 2011). The primary emphasis of this paper is on the evaluation of security concerns pertaining to respiratory medical devices. The challenges pertaining to security and privacy in wireless communication are investigated as well as utilization of rigorous security protocols. In particular, public-key cryptography is examined for its connection to expenditure of computational resources. The implication of device or system misuse is here due to the potential exploitation of resource exhaustion by threat actors.

The US FDA (Wang et al, 2017) investigates the regulatory viewpoints pertaining to the protection of medical devices from cyber threats. To perform a comprehensive inventory of medical devices in a healthcare organization, it is necessary to adopt a spontaneous approach that encompasses a diverse range of devices, allowing for significant insights to be gathered (Brueggemann et al, 2020). The study conducted by Liang and Xue (2010) carried out an extensive investigation on a significant dataset to potentially uncover insights pertaining to the relationship between user behaviour and device vulnerability.

According to the study conducted by (Yeng et al, 2022), human factors could make medical devices vulnerable. Additionally, it will provide an overview of individuals' understanding of certain adversaries' functioning. In context, factors such as psychological, social, and cultural aspects on IS knowledge, attitude and behaviour among healthcare staff. To ensure the confidentiality and integrity of medical records, it is crucial to implement measures that protect against activities like Wi-Fi password guessing, ARP poisoning, and reverse engineering.

## 2.2 Security Measures of Pulmonary Healthcare Devices

The emergence of sensors, miniaturized processors, body area networks, and wireless data transmission technologies has materialized the ideas of *ubiquitous* and *pervasive* human wellbeing monitoring, specifically in the context of secure monitoring of pulmonary healthcare devices. This enables the evaluation of physical, physiological, and biochemical parameters in diverse settings without any limitations on activity (Greiwe and Nyenhuis, 2020).

Exploiting users' lack of sensitization is a common tactic employed by cyber criminals, particularly during crisis emergencies, posing a potential risk such as the case of COVID-19 global pandemic. Jessica Wilkerson (2022), a cyber policy advisor at the FDA's Center for Devices and Radiological Health, states that the growing interconnectedness of devices renders them more susceptible to cybersecurity threats than nonconnected devices. Wilkerson emphasizes that compliance with quality system regulations is mandatory for medical device manufacturers, as it encompasses the need to mitigate cybersecurity risks.

## 2.3 Related Work on Pulmonary Devices

Previous research focused on data protection of pulmonary devices with a focus on encryption and related techniques. A major challenge lies in obtaining precise medical data while ensuring the highest level of safety/security for device usage and accompanying data storage, especially when data includes PII. IoT is commonly referred to when discussing healthcare technology and its emerging integrations

- IoT provides the capability to oversee and regulate pre-existing medical conditions.
- Furthermore, it serves to enhance other technologies, especially by utilizing real-time sensors and high-response medical equipment.
- Users suffer a deterioration in their physical well-being. The ultimate advantage is exemplified by practical and advantageous applications that cater to a diverse patient population. The significant global socioeconomic impact in home care makes the development of applications for self-care as a preventative measure for respiratory disorders highly promising.

Two studies specifically address vulnerabilities in wireless communication of pulmonary healthcare devices. The development of IoT in healthcare can be attributed to the progress made in low-power networked systems and medical sensors in recent years. These IoT devices hold the potential to significantly enhance and broaden the level of care in diverse settings and for various demographic groups. As an illustration, initial system prototypes have showcased the capability of IoT to facilitate early identification of clinical deterioration by means of real-time patient monitoring within hospital settings (Ko et al, 2010), (Chipara et al, 2009), (Lim, 2010).

There has been a profound shift in the operational mechanisms of technologies in recent times, largely influenced by the escalating flow of communication and information. Consequently, this has led to substantial advancements in patient treatment, facilitating the delivery of exceptional healthcare services. During the period preceding the emergence of ICT, conventional healthcare systems operated as the main channel for providing healthcare services (Kalid et al, 2018).

Hypertension, diabetes, arthritis, asthma, cancer, Chronic Obstructive Pulmonary Disease (COPD), dementia, and pain, among various other conditions, affect millions of people globally. It is essential to emphasize that timely identification of a chronic disease allows healthcare professionals to deliver necessary treatment before possible complications arise, resulting in cost reduction (Islam and Yuce, 2016).

## 3. Investigations on Vulnerabilities of Smart Inhalers

### 3.1 Procedures for Establishing the Testing Environment

- Devices: The Propeller Health System Smart Inhaler, Teva Digihaler, BreatherFit, and Lookee 02 Ring are among the notable options available in the market
- Data Connectivity: The establishment of a resilient network infrastructure plays a crucial role in facilitating seamless communication between smart devices and their respective applications. This encompasses the monitoring of both Wi-Fi connections and cellular integrations, facilitating the identification of potential vulnerabilities or unauthorized entry attempts.
- Application Proposal for the Monitoring and Analysis of Breath: The development of this application required careful measures to accurately monitor human breath rates and thoroughly analyse any anomaly. This application's functionality may contain support for multi-user scenarios and seamless integration of data from diverse devices.
- Breath Rate Monitoring: Users' breath rate is constantly monitored by the application. The system is designed to detect anomalies such as irregular patterns of respiration.
- Multi-User Support: With the deliberate intent of preserving data integrity and security across different user profiles, the application has been meticulously developed to operate in scenarios involving concurrent connections of multiple users.
- Data Analysis: The application enables the accomplishment of real-time data analysis. Its purpose revolves around serving as a notification system, providing timely updates to users and healthcare practitioners regarding the latest information that could signify potential health concerns.

### 3.2 Identified Vulnerabilities in the Realm of Smart Medical Devices

This section presents an overview of the prominent vulnerabilities observed in digital medical devices. The previously mentioned items comprise:

- Vulnerabilities in Data Transmission: Insufficient internet encryption techniques during transferring data may cause unauthorized access from an adversary. Healthcare is an important fishing pool for hackers. (Alsubaei et al, 2017), reported in a recent technical document that equip a hospital bed with nearly 10 to 15 networked medical devices (Bates, 2020). These devices transmit data to a closed range smartphone through local and cloud penetration for real time analysis.
- Vulnerabilities in the firmware: Firmware that is outdated or lacks security measures can serve as an entry point for hackers to exploit the functionality of a device. Firmware attack modification is volatile because it plays a huge significance to the devices that control the hardware.
- Weaknesses in the Network: Inadequate network security measures can cause intrusions that compromise multiple devices. To ensure the effective performance of network medical devices, it is necessary to use technologies that incorporate both wired and short-range wireless standards. This explains the purpose and utilization of various communication protocols, including ZigBee, BLE, and Wi-Fi.

Vulnerabilities in User Interfaces: Inadequate user authentication methods possess the possibility of leading to unauthorized access to device settings and data. To intercept some challenges faced by healthcare medical devices, we need to comprehend communication structure used by users to authenticate security safety.

### 3.3 Device Selection

The application of medical devices in the context of asthma has been known to translate into advanced quality care purposely designed to improve health outcomes, and potential financial reduction. Patient applications, electronic monitoring devices, and fully integrated digital inhalers are among the digital tools that have been adopted since outbreak of COVID-19 pandemic, this has highlighted the need for strong medical solutions as a means of providing quality patient care for those suffering from asthma.

Consequently, there is a growing number of companies that offer or promote smart medical devices related to asthma. While selecting healthcare pulmonary devices, a thorough evaluation is undertaken, considering numerous factors, to arrive at an informed decision that is effectively communicated through comprehensive analysis.

#### 3.3.1 Lookee O2

The Lookee O2 Oximeter provides individuals with a convenient and accurate means of monitoring their blood oxygen levels and heart rate comfortably. The central emphasis of security concerns lies in the protection of data privacy and the preservation of device integrity. Through vulnerability analysis, we extensively examined numerous case studies in our study on Pulmonary Health Devices.

*Testing vulnerabilities and threats*

- Data Privacy: The main objective of Lookee O2 is to gather detailed health data, which, if accessed without authorization, could potentially compromise privacy. An example that illustrates this phenomenon is the interception of unencrypted data transmission by individuals with malicious intentions.
- Device Integrity: The device is susceptible to serious threats from firmware attacks, which have the capability to activate malicious software and cause disruption or complete loss of functionality. (Muheidat and Tawalbeh, 2021) articulate the imperative of incorporating comprehensive safeguards for IoT infrastructures. The increase in cybersecurity risks and access to malicious applications to IoT systems' sensitive data can be attributed to unconscious use, neglecting password changes, and failure to update devices.
- The implementation of inappropriate security practices escalates the likelihood of a data breach and various other threats. Despite the implementation of various security mechanisms to fortify IoT devices against cyber-attacks, there is a deficiency in the appropriate documentation of security guidelines (Watson, 2018). Implementing end–end encryption of data policy would be the most reliable policy to minimize certain risks.

*3.3.2 Teva Digihaler*

The Teva Digihaler enables the seamless transmission of data to a mobile application for comprehensive documentation of medication usage.

*Testing for Vulnerability and Threat*

- Breach of access: The user's health data can be compromised by hackers through the exploitation of weak authentication protocols. The current situation, which is advantageous for the implementation of smart inhalers, is characterized by the extensive use of technology by patients and the support from taxpayers (Ferrante et al, 2021), the acknowledgment of the significance held by policy organizations.
- Data Security: The manipulation of drug prescription by adversaries can lead to the improper administration of medication. There is a growing emphasis among researchers on the difficulties posed by privacy breaches in IoT settings, particularly in relation to the disclosure of user information such as data, location, and patient data usage.

*3.3.3 Breather Fit*

The Breather Fit device's goal centres around the enhancement of respiratory muscle strength, highlighting ongoing security concerns that pose a significant challenge to ensuring data access and integrity.

*Threats and Vulnerabilities*

- Breach of Data: The training data stored on the device may be susceptible to access by unauthorized individuals. According to the findings of the survey conducted by (Sicari, 2015), it was discovered that health devices commonly store data in plaintext, making them susceptible to unauthorized access. The utmost importance lies in guaranteeing the encryption of stored data and the implementation of robust access controls.
- Data Manipulation: The transmission of inaccurate data to healthcare providers is page limitations. According to a study conducted by Rawat (2024), data breaches have the potential to disseminate manipulated data to healthcare providers, consequently leading to inaccurate treatment plans. To ensure the preservation of data integrity, it is imperative to implement cryptographic hashing and conduct regular integrity checks.

## 4. Case Study: Lookee O2

In this section, a detailed clarification is presented regarding the methodology utilized to establish and evaluate the security characteristics of the Lookee O2 ring, a Bluetooth device that operates autonomously without relying on IP addresses. The methodology incorporates Bluetooth-specific identification techniques and security testing processes to support non-IP-based communication on the device.

### 4.1 Methodical Approach to Testing Outlined

The employed methodology for the identification and testing of this device consists of several crucial stages.

- Device Identification: The Lookee O2 ring operates on Bluetooth technology, eliminating the reliance on an IP address for network communication. Bluetooth is often used for direct, local device identification and interaction, while IP addresses are used for broader, networked identification and communication. In comparison, our reliance was placed on Bluetooth-specific identification techniques. The serial number (SN) and MAC address were utilized for the sole purpose of device identification.

- *Bluetooth Scanning: We utilized Bluetooth scanning tools to identify devices that are in the vicinity. To accomplish this, we leveraged "Bluetooth View" (the display or interface that shows available Bluetooth devices within range, allowing users to connect, pair, and manage those devices). In this enumeration, devices can be accurately identified through their individual Bluetooth MAC addresses.*
- *Serial Number Search: By employing a scanning tool of choice, we conducted a search for the partial serial number, which aligns with a segment of the entire serial number of the Lookee O2 ring. Through this search, the device was successfully identified.*
- *Bluetooth Traffic Analysis: Once we identified the Lookee O2 ring using its MAC address, our subsequent step involved examining Bluetooth traffic to evaluate the device's security.*
- *The tools we utilized include Wireshark with Bluetooth Adapter. The combination of Wireshark and a Bluetooth adapter was utilized to capture and analyse Bluetooth Low Energy (BLE) traffic. Through the utilization of this methodology, we achieved the capability to monitor the data exchanges between the Lookee O2 ring and its paired devices, effectively pinpointing any security issues that may arise during data transmission.*
- *Service Discovery Protocol (SDP) Tools Overview: The employment of sdptool was carried out to execute Service Discovery Protocol (SDP) scans on the Lookee O2 ring. By employing this tool, it became possible to systematically identify and analyse the services and profiles offered by the device, which plays a pivotal role in comprehending its functionality and pinpointing potential vulnerabilities.*
- *Security Testing The security of the Lookee O2 ring was evaluated through a series of tests conducted with specialized Bluetooth security tools.*

### 4.2 Patient's Information on Lookee O2 Ring Via Vi-Health

This report presents a detailed examination of the security protocols implemented for safeguarding patient data in the Vi-Health application, with specific attention given to the data got from the Lookee O2 Ring device. The dataset includes confidential health measurements, such as oxygen saturation and pulse rates. The report provides a thorough analysis of vulnerabilities, risks, and recommendations for safeguarding this data against unauthorized access or breaches.

#### 4.2.1 Data organization and structuring

The data depicted the tracking of patient vital signs, specifically oxygen saturation and pulse rates, that were captured through the employment of the Lookee O2 Ring device. Every record correlates with a particular session during which the patient's vital signs were observed over a designated duration. Following that, the data was sorted into five distinct labels, with each label representing a distinct patient monitoring session. The organization of the labels and the accompanying data is illustrated in Figure 1 through Figure 3.

| Label | Date & Time | Oxygen Level Average | Pulse Rate Average | Measurement Time | Oxygen Level Range | Pulse Rate Range | Oxygen Level Distribution (95%-100%) |
|---|---|---|---|---|---|---|---|
| A | 16 May 2024, 09:42 AM | 98% | 70/min | 32s | 98-98% | 70-71/min | 32s (100%) |
| B | 12 Aug 2024, 08:51 PM - 11:14 PM | 97% | 64/min | 2h 23m 36s | 92-99% | 53-86/min | 2h 22m 51s (99.49%) |
| C | 18 Jul 2024, 09:30 PM - 09:32 PM | 98% | 75/min | 2m 44s | 97-98% | 70-85/min | 2m 44s (100%) |
| D | 31 Jul 2024, 09:47 AM - 09:58 AM | 99% | 70/min | 11m 28s | 98-99% | 62-77/min | 11m 28s (100%) |
| E | 31 Jul 2024, 10:16 AM - 10:29 AM | 98% | 72/min | 13m 28s | 96-99% | 61-102/min | 13m 28s (100%) |

**Figure 1: Patient Data Information - Smart Inhaler**

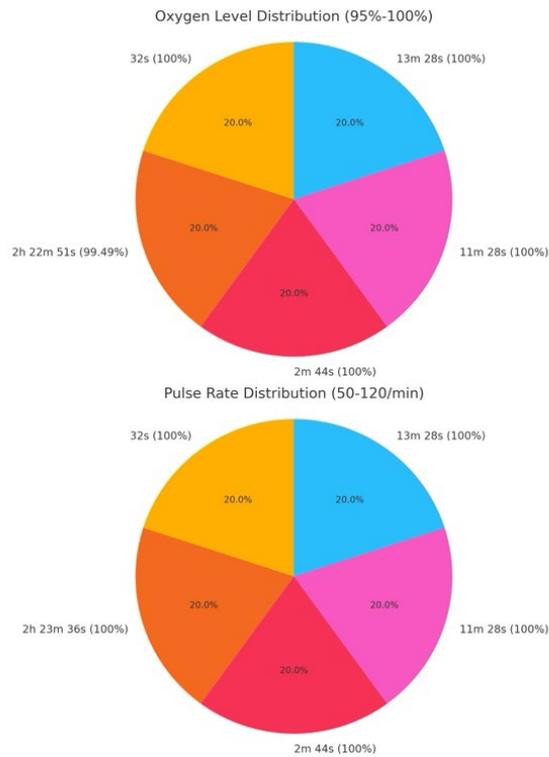

**Figure 2: Oxygen Level and Pulse Rate Distribution**

*4.2.2    Purpose of data organization*

The data was structured in this particular format to optimize analysis, particularly in evaluating the consistency of oxygen levels and pulse rates during different monitoring sessions. The implementation of this framework provides a means for easy comparison and visualization of data, which is of utmost importance for research undertakings, notably in understanding trends, identifying anomalies, and drawing conclusions regarding the efficacy of the Lookee O2 Ring in monitoring patient vitals. The explanation provided clarifies the organization of the data after it was uploaded, highlighting the importance of this arrangement in facilitating thorough research and analysis.

- Data Description: The dataset contains confidential medical data, including oxygen levels and pulse rates, which can infer a patient's health condition.
- Personal Identifiable Information (PII): When linked to a patient's identity, this data becomes highly sensitive and causes the utmost level of protection.

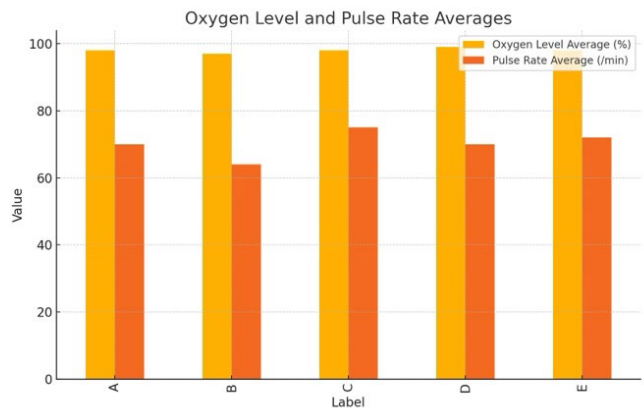

**Figure 3: Oxygen Level and Purse Rate bar c*hart***

### 4.3 Potential Vulnerabilities

- Data Encryption: When storing data on a device or in cloud storage, it is imperative to use robust encryption techniques, such as AES-256, to maintain its security. In Transit: In order to prevent interception, it is crucial to use end-to-end encryption (such as TLS/SSL) when transmitting data between the Lookee O2 Ring, Vi-Health application, and all cloud services.
- Access Control Authorization: Implementing proper role-based access control is essential to limit access to specific data to allow users only.
- Data Breach Risks: Insecure APIs: The prioritization of security measures for external service communication APIs in the Vi-Health application is crucial to minimize the possibility of unauthorized access or exploitation.
- Cloud Security: When data is stored in the cloud, it is essential to confirm that the cloud service provider strictly adheres to rigorous security protocols. This includes conducting routine security audits and implementing suitable safeguards such as encryption and access controls.
- User Education Enhancing User Awareness: Sharing information about recommended practices, such as not using public Wi-Fi for sensitive activities, using strong passwords, and regularly updating software, can effectively lower the risk of unauthorized access or data breaches.

### 4.4 Vulnerability Report

We downloaded and installed applications on the Lookee O2 Ring device to ensure proper functioning. We configured a secure network environment for testing and integrating Wi-Fi and cellular connections. Based on our environmental settings described above, no significant vulnerabilities were identified in the Lookee O2 Ring device. We recommend extending the suggested safeguarding measures to other medical devices as well. Due to page limitations, we will continue to further discuss other devices in our future work.

## 5. Discussion

In this section, we discuss areas for future improvement.

### 5.1 Evaluation of Security Vulnerabilities Through Penetration Testing

- Identify entry points by examining all points of interaction such as Bluetooth, Wi-Fi, and cloud services.
- Perform penetration tests using tools such as Metasploit and Wireshark to detect vulnerabilities in the communication and data storage protocols.
- Simulate attacks by performing network-based exploits such as man-in-the-middle (MITM), denial of service (DoS), and unauthorized access attempts

### 5.2 Network Analysis

- Network analysis tools are used to oversee data packets exchanged between devices and their corresponding applications.
- An assessment of the data flow is conducted to detect any occurrences of unencrypted data and insecure communication channels.
- Conduct an evaluation of encryption by assessing both the robustness and implementation of the encryption protocols. Testing for functionality and performance.

### 5.3 Breath Monitoring

- Employ companion applications to consistently monitor the breath rate data from each device.
- Perform Breathing Simulations: Evaluate different breathing scenarios encompassing regular, erratic, and accelerated respiration patterns.
- Ob Obtain and assess breath rate data for any irregularities

### 5.4 Scenarios in A Multi-User Context

- Implementing multiple user profiles: Creating multiple user accounts within applications.
- The data integrity and performance of the devices are verified by subjecting them to a test involving multiple users.

### 5.5 Proposed Recommendations

- Reinforcing Encryption: Implement resilient and up-to-date encryption protocols to guarantee secure data transmission and storage.
- Strengthening Authentication: Strengthening authentication techniques, such as implementing multifactor authentication.
- User Education: Implementing regular training programs to foster the development of sound security practices among users.
- Providing regular updates in a timely manner is essential to address vulnerabilities in devices and applications.
- Improving Performance: Optimize the application's efficiency in handling real-time data updates and concurrent usage scenarios.

## 6. Conclusion

Due mainly to the sensitivity and risk surrounding user's data, it is critical to maintain confidentiality, integrity, and availability of data privacy for the patient, causing very stringent security protocols for medical data. This shows the inevitable need for cybersecurity measures for healthcare data preservation according to our analysis. We suggest encryption and MFA for both the protection of data and the prevention of hackers. With proper monitoring and constant security checks, it can lead to proper detection of bugs. We firmly believe that these will reduce risks to provide a watertight environment for the sharing of private health information.

To ensure the prevention of unauthorized access to data, it is essential to establish protective measures, with encryption serving as a necessary component. Ensuring strong authentication and authorization mechanisms is crucial for effectively managing data access. Conducting routine security audits is imperative for the detection and mitigation of potential vulnerabilities. To minimize security breaches caused by human error, it is imperative to place the utmost importance on user education. Implementing these suggestions can fortify the security protocols of a user application, reducing the likelihood of data breaches and ensuring the protection of sensitive patient data.